\documentclass[paper]{JHEP3} 

\JHEPspecialurl{http://jhep.sissa.it/JOURNAL/JHEP3.tar.gz}

\usepackage{epsfig,multicol}

\newcommand\fverb{\setbox\pippobox=\hbox\bgroup\verb}
\newcommand\fverbdo{\egroup\medskip\noindent%
			\fbox{\unhbox\pippobox}\ }
\newcommand\fverbit{\egroup\item[\fbox{\unhbox\pippobox}]}
\newcommand {\beq}{\begin{equation}}
\newcommand {\eeq}{\end{equation}}            
\newcommand {\bea}{\begin{eqnarray}}
\newcommand {\eea}{\end{eqnarray}}
\newcommand {\nn}{\nonumber}
\newcommand {\tr}{{\rm tr\,}}

\newcommand {\dd}{\mbox{d}}

\newcommand {\bA}{\tt A}
\newcommand {\bB}{\tt B}
\newcommand {\bC}{\tt C}
\newcommand {\limit}{\rightarrow}

\newbox\pippobox

\title{Super Yang-Mills Theories on the Two-Dimensional Lattice with 
Exact Supersymmetry}

\author{Fumihiko Sugino\\
School of Physics \& BK-21 Physics Division\\
Seoul National University, 
Seoul 151-747, Korea\\
E-mail: \email{sugino@phya.snu.ac.kr}}

\preprint{SNUST-040101 \\\heplat{0401017}}	

\abstract{We construct super Yang-Mills theories 
with ${\cal N}=2, 4$ supersymmetries on the two-dimensional square lattice 
keeping one or two supercharges exactly. 
Along the same line as the previous paper \cite{sugino}, the construction is 
based on topological field theory formulation. 
In order to resolve the problem 
of degenerate classical vacua encountered in the previous paper, 
we consider two kinds of modifications of the action. 
For one of them, any supersymmetry breaking terms do not need to be introduced, and 
the formulations exactly realize some of supersymmetries at the lattice level. 
Our lattice actions flow to the desired continuum theories without any 
fine-tuning of parameters. 
} 


\begin{document} 


\section{Introduction}

Lattice formulations of supersymmetric theories have a long history and 
still have continued to be  
vigorously investigated \cite{nicolai, fujikawa, elitzur, 
kikukawa-nakayama, nishimura, bietenholz, 
ichimatsu, catterall-karamov}\footnote{For a recent review, see \cite{feo}.}. 
Recently, for super Yang-Mills (SYM) theories, 
an interesting approach motivated from 
the idea of deconstruction was presented \cite{kaplan, kaplan2}\footnote{For 
related works, see \cite{kaplan_related, giedt}.}. 
Also, 
various theories without gauge symmetry were discussed based on 
the connection to topological field theory \cite{catterall}. 

In the previous paper \cite{sugino}, we constructed   
SYM theories with extended supersymmetry on the hypercubic 
lattices of various dimensions. 
The construction is based on topological field theory 
formulation \cite{tft, btft}, and 
keeps one or two supercharges manifestly. 
The gauge fields are expressed as ordinary compact unitary variables 
on the lattice links. 
The lattice actions have a number of 
the classical vacua with the degeneracy growing as 
exponential of the number of plaquettes, which makes unclear their 
connection to the perturbative regime of the SYM theories. 
In order to resolve the degeneracy and to 
single out the vacuum corresponding to the target theory, 
we added to the action some supersymmetry breaking term 
which is tuned to vanish in the continuum limit. 

In this paper, the same kind of lattice models are considered for 
${\cal N}=2, \,4$ supersymmetries in two-dimensions. 
We consider to make suitable modifications to the action so that 
the vacuum degeneracy is completely resolved {\em with keeping the exact 
supersymmetry}. 
Two such modifications are presented here. 
One is a modification to 
impose the so-called admissibility conditions on plaquette variables 
similar to 
those for gauge fields coupled to the Ginsparg-Wilson fermions 
\cite{ginsparg-wilson}. It 
will be  
applicable to more general cases not restricted to the SU$(N)$ gauge group. 
It maintains 
the exact supersymmetry possessed by the original 
models in \cite{sugino}, 
and we do not have to add any supersymmetry breaking terms. 
The supersymmetry is exactly realized at the lattice 
level in the SYM theories.
The other is a simple modification of the function $\Phi(x)$ in the lattice actions 
for the gauge group SU$(N)$ as $\Phi(x) \limit \Phi(x) + \Delta\Phi(x)$ with 
the trace parts of some adjoint fields introduced. The actions are supersymmetric, but 
have would-be zero-modes inducing nontrivial constraints among the fields. 
Since such constraints do not appear in the target continuum theories, we should soak up 
the would-be zero-modes to avoid obtaining them. As a result of the insertion of the would-be zero-modes, 
the supersymmetry is violated. 
The supersymmetry breaking effect is much more irrelevant in the continuum limit 
compared to the case in the previous paper \cite{sugino}.  

This paper is organized as follows. In section \ref{sec:review}, we 
briefly review 
the lattice actions constructed in \cite{sugino} for the 
${\cal N}=2, 4$ theories in two-dimensions. For the case of the SU$(N)$ gauge group, 
we explain the structure of the degenerate minima, emphasizing the difference from the U$(N)$ case.  
In section \ref{sec:admissibility}, we impose the 
admissibility conditions on plaquette variables, where the exact 
supersymmetry is preserved by doing a similar modification to the action 
as in ref. \cite{luscher}.  
In section \ref{sec:modification1}, we discuss on the modification 
$\Phi(x) \limit \Phi(x) + \Delta\Phi(x)$ to resolve the problem of 
the degenerate vacua.  
Section \ref{sec:summary} is devoted to the summary and discussions. 

Throughout this paper, we consider the gauge groups $G= \mbox{SU}(N)$, 
$\mbox{U}(N)$ and 
the two-dimensional square lattice.  
Gauge fields are expressed as compact unitary variables 
\beq
U_{\mu}(x)= e^{iaA_{\mu}(x)}
\label{unitary}
\eeq 
on the link $(x, x+\hat{\mu})$. 
`$a$' stands for the lattice spacing, 
and $x\in {\bf Z}^2$ the lattice site. 
All other fields are put on the sites and expanded by a basis of 
$N\times N$ (traceless) hermitian matrices $T^a$. 

\setcounter{equation}{0}
\section{Brief Review of Lattice Actions}
\label{sec:review}

We briefly review the lattice actions of two-dimensional ${\cal N}=2, 4$ 
SYM theories discussed in \cite{sugino}. 

\subsection{${\cal N}= 2$ Case} 
\label{sec:review_N=2}

Other than the gauge variables $U_{\mu}(x)$, 
the ${\cal N}=2$ theory has complex scalars $\phi(x)$, $\bar{\phi}(x)$, 
and fermions are denoted as $\psi_{\mu}(x)$, $\chi(x)$, $\eta(x)$ 
\cite{tft}.   
They are transformed under the exact supersymmetry $Q$ as 
\bea
 & & QU_{\mu}(x) = i\psi_{\mu}(x) U_{\mu}(x), \nn \\
 & & Q\psi_{\mu}(x) = i\psi_{\mu}(x)\psi_{\mu}(x) 
    -i\left(\phi(x) - U_{\mu}(x)\phi(x+\hat{\mu})U_{\mu}(x)^{\dagger}\right),
  \nn \\
 & & Q\phi(x) = 0,     \nn \\
 & & Q\chi(x) = H(x), \quad 
           QH(x) = [\phi(x), \,\chi(x)], \nn \\
 & & Q\bar{\phi}(x) = \eta(x), \quad  Q\eta(x) = [\phi(x), \,\bar{\phi}(x)],  
\label{Q_lattice}
\eea
where $H(x)$ is an auxiliary fields. 
$Q$ is nilpotent up to an 
infinitesimal gauge transformation with the parameter 
$\phi(x)$. In the expansion 
$H(x) = \sum_a H^a(x)T^a$, coefficients 
$H^a(x)$ are real. 
$\phi^a(x)$, $\bar{\phi}^a(x)$ are complex, and 
the fermionic variables $\psi_{\mu}^a(x)$, $\chi^a(x)$, $\eta^a(x)$ 
may be regarded as complexified Grassmann\footnote{A complexified
Grassmann number takes the form: 
$(\mbox{complex number})\times (\mbox{real Grassmann number})$.} 
to be compatible to the U$(1)_R$ rotations (\ref{U1R}). 
Notice that in the path integral $\phi^a(x)$ and $\bar{\phi}^a(x)$ can 
be treated as independent variables and that each of $H^a(x)$ is allowed to be 
shifted by a complex number. Thus, (\ref{Q_lattice}) is consistently 
closed in the path integral expression of the theory.  

The lattice action is  
\bea
S^{{\rm LAT}}_{{\cal N}=2} & = & Q\frac{1}{2g_0^2}\sum_x \, \tr\left[ 
\frac14 \eta(x)\, [\phi(x), \,\bar{\phi}(x)] -i\chi(x)\Phi(x) 
+\chi(x)H(x)\right. \nn \\
 & & \hspace{2cm}\left. \frac{}{} 
+i\sum_{\mu=1}^2\psi_{\mu}(x)\left(\bar{\phi}(x) - 
U_{\mu}(x)\bar{\phi}(x+\hat{\mu})U_{\mu}(x)^{\dagger}\right)\right], 
\label{lat_N=2_S}
\eea
where 
\beq
\Phi(x) =  -i\left[U_{12}(x)- U_{21}(x)\right],   
\label{Phi_2d} 
\eeq
$U_{\mu\nu}(x)$ are plaquette variables written as
\beq
U_{\mu\nu}(x) \equiv U_{\mu}(x) U_{\nu}(x+\hat{\mu}) 
U_{\mu}(x+\hat{\nu})^{\dagger} U_{\nu}(x)^{\dagger}. 
\eeq
The action (\ref{lat_N=2_S}) is clearly $Q$-invariant 
from its $Q$-exact form. 
Also, the invariance under the following global U$(1)_R$ rotation 
holds: 
\bea
U_{\mu}(x) \limit U_{\mu}(x), & & 
\psi_{\mu}(x) \limit e^{i\alpha}\psi_{\mu}(x), \nn \\ 
\phi(x) \limit e^{2i\alpha}\phi(x), &  &  \nn \\
\chi(x) \limit e^{-i\alpha}\chi(x), & & H(x) \limit H(x), \nn \\
\bar{\phi}(x) \limit e^{-2i\alpha}\bar{\phi}(x), & & 
\eta(x) \limit e^{-i\alpha}\eta(x). 
\label{U1R}
\eea
The U$(1)_R$ charge of each variable is read off from (\ref{U1R}), and 
$Q$ increases the charge by one. 
 
After acting $Q$ in the RHS of (\ref{lat_N=2_S}), the action is expressed as  
\bea
S^{{\rm LAT}}_{{\cal N}=2} & = & \frac{1}{2g_0^2}\sum_x \, \tr\left[
\frac14 [\phi(x), \,\bar{\phi}(x)]^2 + H(x)^2 
-iH(x)\Phi(x) \right. \nn \\
 & & \hspace{1.5cm}
+\sum_{\mu=1}^2\left(\phi(x)-U_{\mu}(x)\phi(x+\hat{\mu})U_{\mu}(x)^{\dagger}
\right)\left(\bar{\phi}(x)-U_{\mu}(x)\bar{\phi}(x+\hat{\mu})
U_{\mu}(x)^{\dagger}\right) \nn \\
 & & \hspace{1.5cm} -\frac14 \eta(x)[\phi(x), \,\eta(x)] 
- \chi(x)[\phi(x), \,\chi(x)] \nn \\
 & & \hspace{1.5cm}
-\sum_{\mu=1}^2\psi_{\mu}(x)\psi_{\mu}(x)\left(\bar{\phi}(x)  + 
U_{\mu}(x)\bar{\phi}(x+\hat{\mu})U_{\mu}(x)^{\dagger}\right) \nn \\
 & & \hspace{1.5cm}\left. \frac{}{}+ i\chi(x) Q\Phi(x) 
-i\sum_{\mu=1}^2\psi_{\mu}(x)\left(\eta(x)-
U_{\mu}(x)\eta(x+\hat{\mu})U_{\mu}(x)^{\dagger}\right)\right]. 
\label{lat_N=2_S2}
\eea
Integration of $H(x)$ induces $\Phi(x)^2$ term, which    
yields the gauge kinetic term as the form 
\beq
\frac{1}{8g_0^2}\sum_x
\tr\left[-(U_{12}(x) - U_{21}(x))^2\right],  
\label{gauge_kin}
\eeq
which leads a number of the classical vacua  
\beq
U_{12}(x) = \left( \begin{array}{ccc} \pm 1 &        &       \\   
                                                & \ddots &       \\
                                                &        & \pm 1
\end{array}\right) 
\label{huge_minima}
\eeq
up to gauge transformations for $G=\mbox{U}(N)$, 
where any combinations of $\pm 1$ 
are allowed in the diagonal entries. 
Since an arbitrary configuration of (\ref{huge_minima}) can be 
taken for each plaquette, it leads the degeneracy 
growing as exponential of the number of plaquettes. 
In order to investigate the dynamics of the model, 
we need to sum up contributions 
from all of the minima,  
and the ordinary weak field 
expansion around a single vacuum 
$U_{12}(x)=1$ can not be justified. 

\paragraph{Degenerate Minima for $G=\mbox{SU}(N)$}

For the case $G=\mbox{SU}(N)$, we have more complicated structures of the 
degenerate minima. Notice that because the adjoint fields are {\em traceless}, 
the trace part of $\Phi(x)$ does not appear in the expression (\ref{lat_N=2_S2})\footnote{
We thank Y. Kikukawa for pointing out this issue.}. 
The minima are given by the configurations satisfying 
\beq
\Phi(x) - \left(\frac{1}{N}\tr \Phi(x)\right){\bf 1}_N =0 
\label{minima_SU(N)}
\eeq
instead of $\Phi(x)=0$. 
The solutions of (\ref{minima_SU(N)}) are classified to the following three types:
\begin{enumerate}
\item 
the same form as in the case $G=\mbox{U}(N)$ 
\beq
U_{12}(x) = \left( \begin{array}{ccc} \pm 1 &        &       \\   
                                                & \ddots &       \\
                                                &        & \pm 1
\end{array}\right) 
\label{SU(N)minima1}
\eeq
with `$-1$' appearing even times 
in the diagonal entries

\item 
the SU$(N)$ center 
\beq
U_{12}(x) = e^{i\frac{2\pi}{N}k(x)}\, {\bf 1}_N \, \in Z_N
\qquad (k(x) = 0, \cdots, N-1)
\label{SU(N)minima2}
\eeq

\item
the solution appearing only when $N=4n$ ($n=1, 2, \cdots$)
\beq
U_{12}(x) = 
\left( \begin{array}{cccccc} e^{i\alpha(x)} &        &                &                  &     &    \\   
                                            & \ddots &                &                  &     &     \\
                                            &        & e^{i\alpha(x)} &                 &    &    \\
                                            &        &                & -e^{-i\alpha(x)} &       &    \\
                                            &        &                &                  & \ddots &    \\
                                            &        &                &                  &        & -e^{-i\alpha(x)}     
\end{array}\right) 
\label{SU(N)minima3}
\eeq
up to gauge transformations. In the diagonal entries, both of $e^{i\alpha(x)}$ and $-e^{-i\alpha(x)}$ appear 
$2n$ times. $\alpha(x)$ is an arbitrary phase.  
\end{enumerate}  

Again, due to the huge number of classical minima, 
the ordinary weak field 
expansion around a single vacuum 
$U_{12}(x)=1$ can not be justified either in the SU$(N)$ case.   

\subsection{${\cal N}= 4$ Case} 
\label{sec:review_N=4}

In the ${\cal N}=4$ theory, scalar fields $B(x)$, $C(x)$ appear 
as well as $\phi(x)$ and $\bar{\phi}(x)$. 
Auxiliary fields $\tilde{H}_{\mu}(x)$, $H(x)$ are introduced, and 
fermions are $\psi_{\pm}(x)$, $\chi_{\pm}(x)$, $\eta_{\pm}(x)$ 
\cite{btft}\footnote{$B^a(x)$, $C^a(x)$, $\tilde{H}_{\mu}^a(x)$, 
$H^a(x)$ are real, and $\psi_{\pm}^a(x)$, $\chi_{\pm}^a(x)$, 
$\eta_{\pm}^a(x)$ may be regarded as complexified Grassmann.}. 
The latticization keeps two supercharges $Q_{\pm}$, which transform 
the fields as 
\bea
 & & Q_+U_{\mu}(x) = i\psi_{+\mu}(x)U_{\mu}(x), \nn \\
 & & Q_-U_{\mu}(x) = i\psi_{-\mu}(x)U_{\mu}(x), \nn \\
 & & Q_+\psi_{+\mu}(x) = i\psi_{+\mu}(x) \psi_{+\mu}(x) 
  -i\left(\phi(x)-U_{\mu}(x)\phi(x+\hat{\mu})U_{\mu}(x)^{\dagger}\right), 
\nn \\
 & & Q_-\psi_{-\mu}(x) = i\psi_{-\mu}(x) \psi_{-\mu}(x) 
  +i\left(\bar{\phi}(x)-U_{\mu}(x)\bar{\phi}(x+\hat{\mu})U_{\mu}(x)^{\dagger}
\right), \nn \\
 & & Q_-\psi_{+\mu}(x) = \frac{i}{2}
\left\{\psi_{+\mu}(x), \,\psi_{-\mu}(x)\right\} -\frac{i}{2}
\left(C(x)-U_{\mu}(x)C(x+\hat{\mu})U_{\mu}(x)^{\dagger}\right) 
-\tilde{H}_{\mu}(x), \nn \\
 & & Q_+\psi_{-\mu}(x) = \frac{i}{2}
\left\{\psi_{+\mu}(x), \,\psi_{-\mu}(x)\right\} -\frac{i}{2}
\left(C(x)-U_{\mu}(x)C(x+\hat{\mu})U_{\mu}(x)^{\dagger}\right) 
+\tilde{H}_{\mu}(x), \nn \\
 & & Q_+\tilde{H}_{\mu}(x) = -\frac12
\left[\psi_{-\mu}(x), \,\phi(x)+U_{\mu}(x)\phi(x+\hat{\mu})U_{\mu}(x)^{\dagger}
\right] \nn \\
 & & \hspace{2cm} 
+\frac14\left[\psi_{+\mu}(x), \, C(x) +U_{\mu}(x)C(x+\hat{\mu})
U_{\mu}(x)^{\dagger}\right] \nn \\
 & & \hspace{2cm} +\frac{i}{2}\left(\eta_+(x) 
-U_{\mu}(x)\eta_+(x+\hat{\mu})U_{\mu}(x)^{\dagger}\right) \nn \\
 & & \hspace{2cm}  
+\frac{i}{2}\left[\psi_{+\mu}(x), \,\tilde{H}_{\mu}(x)\right] 
+\frac14\left[\psi_{+\mu}(x)\psi_{+\mu}(x), \,\psi_{-\mu}(x)\right], \nn \\
 & & Q_-\tilde{H}_{\mu}(x) = -\frac12
\left[\psi_{+\mu}(x), \,\bar{\phi}(x)+U_{\mu}(x)\bar{\phi}(x+\hat{\mu})
U_{\mu}(x)^{\dagger}\right] \nn \\
 & & \hspace{2cm}
-\frac14\left[\psi_{-\mu}(x), \, C(x) +U_{\mu}(x)C(x+\hat{\mu})
U_{\mu}(x)^{\dagger}\right] \nn \\
 & & \hspace{2cm} -\frac{i}{2}\left(\eta_-(x) 
-U_{\mu}(x)\eta_-(x+\hat{\mu})U_{\mu}(x)^{\dagger}\right) \nn \\
 & & \hspace{2cm}
+\frac{i}{2}\left[\psi_{-\mu}(x), \,\tilde{H}_{\mu}(x)\right] 
-\frac14\left[\psi_{-\mu}(x)\psi_{-\mu}(x), \,\psi_{+\mu}(x)\right],  
\label{group_A_lat}
\eea
\bea
 & & Q_+B(x) = \chi_+(x), \quad Q_+\chi_+(x) = [\phi(x), \,B(x)], \quad 
Q_-\chi_+(x) = \frac12[C(x), \,B(x)]-H(x), \nn \\
 & & Q_-B(x) = \chi_-(x), \quad Q_-\chi_-(x) = -[\bar{\phi}(x), \,B(x)], \quad 
Q_+\chi_-(x) = \frac12[C(x), \,B(x)]+H(x), \nn \\
 & & Q_+H(x) = [\phi(x), \,\chi_-(x)] +\frac12[B(x), \,\eta_+(x)] 
-\frac12[C(x), \,\chi_+(x)],  \nn \\
 & &  Q_-H(x) = [\bar{\phi}(x), \,\chi_+(x)] -\frac12[B(x), \,\eta_-(x)] 
+\frac12[C(x), \,\chi_-(x)], 
\label{group_B_2d}
\eea
\bea
 & & Q_+C(x) = \eta_+(x), \quad Q_+\eta_+(x) = [\phi(x), \,C(x)], \quad 
Q_-\eta_+(x) = -[\phi(x), \,\bar{\phi}(x)], \nn \\
 & & Q_-C(x) = \eta_-(x), \quad Q_-\eta_-(x) = -[\bar{\phi}(x), \,C(x)], \quad 
Q_+\eta_-(x) = [\phi(x), \,\bar{\phi}(x)], \nn \\
 & & Q_+\phi(x) = 0, \quad Q_-\phi(x)= -\eta_+(x), \quad 
Q_+\bar{\phi}(x) = \eta_-(x), \quad Q_-\bar{\phi}(x) = 0.  
\label{group_C}
\eea
The transformation leads the following nilpotency of $Q_{\pm}$ 
(up to gauge transformations): 
\bea
Q_+^2 & = & 
(\mbox{infinitesimal gauge transformation with the parameter }\phi), \nn \\
Q_-^2 & = & 
(\mbox{infinitesimal gauge transformation with the parameter }-\bar{\phi}), 
\nn \\
\{Q_+, Q_-\} & = & 
 (\mbox{infinitesimal gauge transformation with the parameter }C). 
\label{nilpotent_N=4}
\eea
  
We constructed the lattice action with exact $Q_{\pm}$ symmetry as 
\bea
S^{{\rm LAT}}_{{\cal N}=4} &  = & 
Q_+Q_-\frac{1}{2g_0^2}\sum_x\, \tr \left[-iB(x)\Phi(x) - 
\sum_{\mu=1}^2\psi_{+\mu}(x)\psi_{-\mu}(x)-\chi_+(x)\chi_-(x) \right. \nn \\ 
 & & \hspace{2.5cm} \left. -\frac14\eta_+(x)\eta_-(x)\right], 
\label{lat_N=4_S_2d}
\eea
where $\Phi(x)$ is given by (\ref{Phi_2d}). 
The action is invariant under the SU$(2)_R$ transformation, 
whose generators are expressed as 
\bea
J_{++} & = & \sum_{x,\, a} \left[
\sum_{\mu}\psi_{+\mu}^a(x)\frac{\partial}{\partial\psi_{-\mu}^a(x)} + 
\chi_{+}^a(x)\frac{\partial}{\partial\chi_{-}^a(x)} - 
\eta_{+}^a(x)\frac{\partial}{\partial\eta_{-}^a(x)} 
+ 2\phi^a(x)\frac{\partial}{\partial C^a(x)} \right. \nn \\
 & & \hspace{1cm} \left. 
- C^a(x)\frac{\partial}{\partial \bar{\phi}^a(x)}\right],  \nn \\
J_{--} & = & \sum_{x, \,a} \left[
\sum_{\mu}\psi_{-\mu}^a(x)\frac{\partial}{\partial\psi_{+\mu}^a(x)} + 
\chi_{-}^a(x)\frac{\partial}{\partial\chi_{+}^a(x)} - 
\eta_{-}^a(x)\frac{\partial}{\partial\eta_{+}^a(x)} 
- 2\bar{\phi}^a(x)\frac{\partial}{\partial C^a(x)} \right. \nn \\ 
 & & \hspace{1cm} \left. 
+ C^a(x)\frac{\partial}{\partial \phi^a(x)}\right],  \nn \\
J_0 & = & \sum_{x, \,a} \left[
\sum_{\mu}\psi_{+\mu}^a(x)\frac{\partial}{\partial\psi_{+\mu}^a(x)} 
-\sum_{\mu}\psi_{-\mu}^a(x)\frac{\partial}{\partial\psi_{-\mu}^a(x)} 
+\chi_{+}^a(x)\frac{\partial}{\partial\chi_{+}^a(x)} 
-\chi_{-}^a(x)\frac{\partial}{\partial\chi_{-}^a(x)} 
\right. \nn \\
 &  & \hspace{1cm} \left. +\eta_{+}^a(x)\frac{\partial}{\partial\eta_{+}^a(x)}
-\eta_{-}^a(x)\frac{\partial}{\partial\eta_{-}^a(x)}  
+2\phi^a(x)\frac{\partial}{\partial\phi^a(x)} 
-2\bar{\phi}^a(x)\frac{\partial}{\partial\bar{\phi}^a(x)}
\right], 
\label{SU2R}
\eea
with the script $a$ being the index of a basis of the gauge group 
generators.  
They form the SU$(2)$ algebra: 
\beq
[J_0, \, J_{++}] = 2J_{++}, \quad [J_0, \, J_{--}] = -2J_{--}, \quad 
[J_{++}, \, J_{--}] = J_0. 
\eeq 
Under the SU$(2)_R$, each of $(\psi^a_{+\mu}, \psi^a_{-\mu})$, 
$(\chi^a_+, \chi^a_-)$, 
$(\eta^a_+, -\eta^a_-)$ and $(Q_+, Q_-)$ transforms as a doublet, 
and $(\phi^a, C^a, -\bar{\phi}^a)$ as a triplet. 
We can easily see the SU$(2)_R$ invariance of the action (\ref{lat_N=4_S_2d}), 
because $Q_+Q_-$, $\tr(\psi_{+\mu}(x)\psi_{-\mu}(x))$, 
$\tr(\chi_+(x)\chi_-(x))$ and $\tr(\eta_+(x)\eta_-(x))$ are SU$(2)_R$ 
singlets. 
Also, the action has the symmetry of exchanging the two supercharges 
$Q_+ \leftrightarrow Q_-$  
with 
\bea
 & & \phi \rightarrow -\bar{\phi}, \quad \bar{\phi} \rightarrow -\phi, \quad 
B \rightarrow -B, \nn \\
 & & \chi_+ \rightarrow -\chi_-, \quad \chi_- \rightarrow -\chi_+, \quad 
\tilde{H}_{\mu} \rightarrow -\tilde{H}_{\mu}, \nn \\
 & & \psi_{\pm\mu} \rightarrow \psi_{\mp\mu}, \quad 
\eta_{\pm} \rightarrow \eta_{\mp}. 
\label{Q_exchange_2d}
\eea

After acting $Q_{\pm}$, the action is more explicitly written as 
\bea
S^{{\rm LAT}}_{{\cal N}=4} & = & \frac{1}{2g_0^2}\sum_x \tr \left[
-i\left(\frac12[C(x), \,B(x)]+H(x)\right)\Phi(x) + H(x)^2 \right. \nn \\
 & & \hspace{2cm} +i\chi_-(x)Q_+\Phi(x) -i\chi_+(x)Q_
-\Phi(x) -iB(x)Q_+Q_-\Phi(x) 
\nn \\
 & & \hspace{2cm} -[\phi(x),\,B(x)][\bar{\phi}(x), \,B(x)]
-\frac14[C(x), \,B(x)]^2 \nn \\
 & & \hspace{2cm} +\chi_+(x)[\bar{\phi}(x), \,\chi_+(x)]
-\chi_-(x)[\phi(x), \, \chi_-(x)] + \chi_-(x) [C(x), \, \chi_+(x)] \nn \\
 & & \hspace{2cm} -\chi_-(x)[B(x), \, \eta_+(x)] 
-\chi_+(x)[B(x), \, \eta_-(x)] \nn \\
 & & \hspace{2cm} -\frac14[\phi(x), \,\bar{\phi}(x)]^2 
-\frac14[\phi(x), \, C(x)][\bar{\phi}(x), \, C(x)] \nn \\
 & & \hspace{2cm} \left. -\frac14\eta_-(x)[\phi(x), \, \eta_-(x)] 
+\frac14\eta_+(x)[\bar{\phi}(x), \, \eta_+(x)] 
-\frac14\eta_+(x)[C(x), \, \eta_-(x)]\right] \nn \\
 & & +\frac{1}{2g_0^2}\sum_{x}\sum_{\mu=1}^2\tr\left[
\tilde{H}_{\mu}(x)^2 
-\frac12\psi_{+\mu}(x)\psi_{+\mu}(x)\psi_{-\mu}(x)\psi_{-\mu}(x) \right. \nn \\
 & & \hspace{2cm} 
+\left(\phi(x) - U_{\mu}(x)\phi(x+\hat{\mu})U_{\mu}(x)^{\dagger}\right)
\left(\bar{\phi}(x) - U_{\mu}(x)\bar{\phi}(x+\hat{\mu})U_{\mu}(x)^{\dagger}
\right)
\nn \\
 & & \hspace{2cm}
+\frac14\left(C(x)-U_{\mu}(x)C(x+\hat{\mu})U_{\mu}(x)^{\dagger}\right)^2 
\nn \\
 & & \hspace{2cm} 
-\psi_{+\mu}(x)\psi_{+\mu}(x)\left(\bar{\phi}(x)+
U_{\mu}(x)\bar{\phi}(x+\hat{\mu})U_{\mu}(x)^{\dagger}\right) \nn \\
 & & \hspace{2cm} 
+\psi_{-\mu}(x)\psi_{-\mu}(x)\left(\phi(x)+
U_{\mu}(x)\phi(x+\hat{\mu})U_{\mu}(x)^{\dagger}\right) \nn \\
 & & \hspace{2cm}-i\psi_{+\mu}(x)\left(\eta_-(x)-
U_{\mu}(x)\eta_-(x+\hat{\mu})U_{\mu}(x)^{\dagger}\right) \nn \\
 & & \hspace{2cm}-i\psi_{-\mu}(x)\left(\eta_+(x)-
U_{\mu}(x)\eta_+(x+\hat{\mu})U_{\mu}(x)^{\dagger}\right) \nn \\
 & & \hspace{2cm}\left. 
-\frac12\left\{\psi_{+\mu}(x), \,\psi_{-\mu}(x)\right\}
\left(C(x) + U_{\mu}(x)C(x+\hat{\mu})U_{\mu}(x)^{\dagger}\right)\right]. 
\label{N=4_lat_S2}
\eea
Similarly to the ${\cal N}=2$ case, the gauge kinetic term, induced 
after integrating $H(x)$ out, suffers from the problem of 
degenerate classical vacua. 

In the previous paper \cite{sugino}, 
for both cases of ${\cal N}=2, \, 4$ theories in two-dimensional 
$M\times M$ periodic lattice, we added the term 
\beq
\frac{1}{2g_0^2}\, \rho \sum_x\tr 
\left(2 - U_{12}(x) - U_{21}(x)\right)
\label{SUSY_breaking}
\eeq
to the action with $\rho= \frac{1}{M^s}$ ($0<s<2$) to resolve the vacuum degeneracy. 
This term breaks the exact supersymmetries $Q$ and $Q_{\pm}$
respectively, 
but it is tuned to vanish in the continuum limit. 
In what follows, 
we will consider to resolve the degeneracy keeping the exact 
supersymmetries without introducing the breaking term (\ref{SUSY_breaking}).

\setcounter{equation}{0}
\section{Admissibility Conditions}
\label{sec:admissibility}

In order to single out the vacuum $U_{12}(x)=1$ from the degeneracy, 
we impose the so-called admissibility condition 
\beq
||1-U_{12}(x)|| < \epsilon  
\label{admissibility}
\eeq  
on each plaquette variable. 
For definiteness, we use the following 
definition of the norm of a matrix $A$:   
\beq
||A|| \equiv \left[\tr \left(AA^{\dagger}\right)\right]^{1/2}, 
\eeq
and then $\epsilon$ is a positive number chosen as 
$0 < \epsilon < 2$ for $G=\mbox{U}(N)$. 
Also, for $G=\mbox{SU}(N)$ we choose 
\bea
 & & 0 < \epsilon < 2\sqrt{2}   \qquad \hspace{1.7cm} (N=2, 3, 4) \\
 & & 0 < \epsilon < 2\sqrt{N}\sin\left(\frac{\pi}{N}\right)   \qquad (N\geq 5),  
\eea
so that excluded are the minima (\ref{SU(N)minima1}, \ref{SU(N)minima2}, \ref{SU(N)minima3}) 
other than $U_{12}(x) = 1$. 
The same kind of condition was introduced for gauge fields coupled to 
the Ginsparg-Wilson fermions \cite{ginsparg-wilson}.  
Notice that  (\ref{admissibility}) is a gauge invariant statement 
and that $||1-U_{12}(x)||^2$ is proportional to the standard Wilson action: 
\beq
||1-U_{12}(x)||^2 = \tr\left[2-U_{12}(x)-U_{21}(x)\right]. 
\eeq

\subsection{${\cal N}=2$ Case}

We modify the action of the ${\cal N}=2$ theory 
as follows: \\
\noindent 
When $||1-U_{12}(x)|| < \epsilon$ for $\forall x$, 
\bea
\hat{S}^{{\rm LAT}}_{{\cal N}=2} & = & 
 Q\frac{1}{2g_0^2}\sum_x \, \tr\left[ 
\frac14 \eta(x)\, [\phi(x), \,\bar{\phi}(x)] -i\chi(x)\hat{\Phi}(x) 
+\chi(x)H(x)\right. \nn \\
 & & \hspace{2cm}\left. \frac{}{} 
+i\sum_{\mu=1}^2\psi_{\mu}(x)\left(\bar{\phi}(x) - 
U_{\mu}(x)\bar{\phi}(x+\hat{\mu})U_{\mu}(x)^{\dagger}\right)\right], 
\label{lat_N=2_Shat}
\eea
and otherwises 
\beq
\hat{S}^{{\rm LAT}}_{{\cal N}=2} = + \infty.   
\label{lat_N=2_Shat2}
\eeq
Here 
\beq
\hat{\Phi}(x) =  \frac{\Phi(x)}{1-\frac{1}{\epsilon^2}||1-U_{12}(x)||^2}.    
\label{Phihat_2d} 
\eeq
The form of the action is somewhat 
similar to that of U(1) gauge theory constructed by 
L\"{u}scher \cite{luscher}. 
Note that the Boltzmann weight 
$\exp\left[-\hat{S}^{{\rm LAT}}_{{\cal N}=2}\right]$ 
is a product of local factors, which guarantees the locality of 
the theory. 
Also, it is easily seen that the Boltzmann weight is smooth and 
differentiable with respect to lattice variables 
for the region except the boundary 
\beq
||1-U_{12}(x)|| = \epsilon. 
\label{boundary}
\eeq
Let us look closer 
at the smoothness on the boundary.  
For the partition function 
\beq
Z = \int [\dd\, (\mbox{fields})]\,  
\exp\left[-\hat{S}^{{\rm LAT}}_{{\cal N}=2}\right], 
\eeq
after the integration over $H(x)$ and fermions, the relevant parts of 
the Boltzmann weight are evaluated as 
\bea
 & & \sum_{\{c(x)\}}d(\{c(x)\})\prod_x{\cal B}(c(x))\, , \\ 
 & & {\cal B}(c(x)) \equiv 
\left(\frac{1}{1- \frac{1}{\epsilon^2}||1-U_{12}(x)||^2}\right)^{c(x)}
\exp\left[-\frac{1}{8g_0^2}\frac{\tr(2-U_{12}(x)^2-U_{21}(x)^2)}{\left(1- 
\frac{1}{\epsilon^2}||1-U_{12}(x)||^2\right)^2}\right]
\label{relevant_part}
\eea
near the boundary. Here, $c(x)$ takes $N^2-1$ or $2(N^2-1)$ 
for each $x$, and $d(\{c(x)\})$ are irrelevant factors. 
It is smooth and differentiable on the boundary, which is essentially same 
as the fact that the function 
\beq
f(t) = \left \{ \begin{array}{cc} \frac{1}{t^n}e^{-c/t^2} & 
\mbox{ for } t>0 \\ 0 & \mbox{ for } t \leq 0 \end{array} \right. 
\eeq
with $c$ positive constant is smooth and differentiable at $t=0$ for 
$n=0, 1, 2, \cdots$. 
Similarly, for the unnormalized correlation function 
among the finite number of operators ${\cal O}_1, \cdots, {\cal O}_k$: 
\beq
\int [\dd\, (\mbox{fields})]\,  {\cal O}_1 \cdots {\cal O}_k \, 
\exp\left[-\hat{S}^{{\rm LAT}}_{{\cal N}=2}\right], 
\eeq 
the Boltzmann weight is smooth and differentiable on the boundary 
(\ref{boundary}), as long as the operators contain the finite number of 
$H(x)$ for each $x$. (Compared to the case of the partition function, 
all the essential difference is that 
some of the powers $c(x)$ in (\ref{relevant_part}) are  
shifted by finite amount, which does not affect the smoothness.)   
It leads the $Q$ invariance of the Boltzmann weight as the following form: 
\beq
\int [\dd\, (\mbox{fields})]\,  {\cal O}_1 \cdots {\cal O}_k \,\,  
Q\left( \exp\left[-\hat{S}^{{\rm LAT}}_{{\cal N}=2}\right]\right) = 0. 
\label{Q_invariance}
\eeq

If we simply imposed (\ref{admissibility}) just by putting the step functions
\beq
\prod_x \theta\left(\epsilon^2 - ||1-U_{12}(x)||^2\right) 
\eeq
in front of the Boltzmann weight 
$\exp\left[-S^{{\rm LAT}}_{{\cal N}=2}\right]$ with the original 
action (\ref{lat_N=2_S}), 
the supersymmetry $Q$ would be broken due to the contribution 
from the boundary (\ref{boundary}). 
The modification of the action (\ref{lat_N=2_Shat}, \ref{lat_N=2_Shat2}) 
however makes 
the breaking effect completely suppressed, and maintains the supersymmetry.  

\paragraph{No Fermion Doublers and Renormalization}

We may expand the exponential of the link variable  
(\ref{unitary}), and the action (\ref{lat_N=2_Shat}, \ref{lat_N=2_Shat2}) 
leads to the ${\cal N}=2$ SYM theory in  
the classical continuum limit:
\beq
a\limit 0 \quad \mbox{ with }\quad g_2^2 \equiv g_0^2/a^2 \quad\mbox{ fixed}.
\label{ccl}
\eeq 
Note that $\epsilon$ is independent of the lattice spacing $a$.  
Also, the modification to the fermionic part of the action reads   
\bea
\tr \left[i\chi(x)\,Q\hat{\Phi}(x)\right] & = & 
\frac{1}{1-\frac{1}{\epsilon^2}||1-U_{12}(x)||^2}
\,\tr\left[i\chi(x)\,Q\Phi(x)\right] \nn \\
 & & -\frac{\tr\left[i\chi(x)\Phi(x)\right]}{\left(1-\frac{1}{\epsilon^2}
||1-U_{12}(x)||^2\right)^2} \frac{1}{\epsilon^2}
\,\tr\left[QU_{12}(x) + QU_{21}(x)\right], 
\eea
where the second term contributes to gauge-fermion interaction terms of the 
irrelevant order $O(a^8)$ but not to 
fermion kinetic terms. 
(Notice that fermionic variables are rescaled as 
$(\mbox{fermions})\limit a^{3/2}(\mbox{fermions})$ 
when taking the continuum limit.) 
Thus, the modification does not affect the fermion 
kinetic terms, and the absence of fermion doubling is shown as 
in the previous paper \cite{sugino}.  
Indeed, the fermion kinetic terms are expressed as 
\beq
S_f^{(2)} =  \frac{a^4}{2g_0^2}\sum_{x, \mu}\tr\left[
-\frac12\Psi(x)^T\gamma_{\mu}(\Delta_{\mu}+\Delta^*_{\mu})\Psi(x) 
-a\frac12\Psi(x)^TP_{\mu}\Delta_{\mu}\Delta^*_{\mu}\Psi(x)\right], 
\label{wilson_like_N=2}
\eeq
where fermions were combined as 
$\Psi^T = \left( \psi_1, \psi_2, \chi, \frac12\eta \right)$. 
The $\gamma$-matrices and $P_{\mu}$ are given by 
\beq
\gamma_1=-i\left( \begin{array}{cc} 0 & \sigma_1 \\
                                \sigma_1 &   0  \end{array}\right), \qquad 
\gamma_2=i\left( \begin{array}{cc} 0 & \sigma_3 \\
                                \sigma_3 &   0  \end{array}\right), \qquad  
P_1 = \left(\begin{array}{cc} 0 & \sigma_2 \\ 
                                  \sigma_2 & 0    \end{array}\right), \qquad
P_2 = -i\left(\begin{array}{cc} 0 & {\bf 1}_2 \\ 
                                  -{\bf 1}_2 & 0    \end{array}\right)  
\eeq
with $\sigma_i$ ($i=1,2,3$) being Pauli matrices. 
Note that they all anticommute each other: 
\beq
\{\gamma_{\mu}, \gamma_{\nu}\} = -2\delta_{\mu\nu}, \quad 
\{ P_{\mu}, P_{\nu}\} = 2\delta_{\mu\nu}, \quad 
\{ \gamma_{\mu}, P_{\nu}\} = 0. 
\label{anticommute_N=2}
\eeq
$\Delta_{\mu}$ and $\Delta^*_{\mu}$ represent forward and backward difference  
operators respectively: 
\beq
\Delta_{\mu}f(x) \equiv \frac{1}{a}\left(f(x+\hat{\mu})-f(x)\right), \quad 
\Delta^*_{\mu}f(x) \equiv \frac{1}{a}\left(f(x)-f(x-\hat{\mu})\right).
\eeq
The kernel of the kinetic terms (\ref{wilson_like_N=2}) is written in the 
momentum space $-\frac{\pi}{a}\leq q_{\mu} < \frac{\pi}{a}$ as 
\beq
D=\sum_{\mu=1}^2\left[-i\gamma_{\mu}\frac{1}{a}\sin \left(q_{\mu}a\right)
+2P_{\mu}\frac{1}{a}\sin^2 \left(\frac{q_{\mu}a}{2}\right)\right]. 
\label{D-kernel}
\eeq
It is easy to see that the kernel $D$ vanishes only at the origin 
$q_1=q_2=0$, because using (\ref{anticommute_N=2}) we get 
\beq
D^2 = \frac{1}{a^2}\sum_{\mu=1}^2\left[\sin^2\left(q_{\mu}a\right)
+4 \sin^4\left(\frac{q_{\mu}a}{2}\right)\right]. 
\label{D_square}
\eeq
Thus, the fermion kinetic terms contain no fermion doublers. 
 
For the renormalization, we can repeat the argument in \cite{sugino} 
without introducing the supersymmetry breaking term (\ref{SUSY_breaking}). 
Note that the U$(1)_R$ symmetry is kept intact under the modification. 
For the $G=\mbox{SU}(N)$, the gauge symmetry and the U$(1)_R$ invariance allow the operator 
$\tr \phi\bar{\phi}$, while it is forbidden by 
the supersymmetry $Q$. 
For the U$(N)$ case, in addition, we should take into account the operator $\tr H$. 
However, it is prohibited by the reflection symmetry: 
$x\equiv (x_1, x_2) \limit \tilde{x}\equiv (x_2, x_1)$ with 
\bea
(U_1(x), U_2(x)) & \limit & (U_2(\tilde{x}), U_1(\tilde{x})) \nn \\
(\psi_1(x), \psi_2(x)) & \limit & (\psi_2(\tilde{x}), \psi_1(\tilde{x})) \nn \\
(H(x), \chi(x)) & \limit & (-H(\tilde{x}), -\chi(\tilde{x})) \nn \\
(\phi(x), \bar{\phi}(x), \eta(x) ) & \limit &  (\phi(\tilde{x}), \bar{\phi}(\tilde{x}), \eta(\tilde{x})). 
\label{reflection}
\eea
Hence, 
radiative corrections do not generate relevant or marginal operators except 
the identity. 
Our modified lattice action is shown 
to flow to the desired continuum theory  
without any fine-tuning. 


\subsection{${\cal N}=4$ Case}

For the case of ${\cal N}=4$ theory, similar modification is possible: \\
\noindent
When $||1-U_{12}(x)|| < \epsilon$ for $\forall x$, 
\bea
\hat{S}^{{\rm LAT}}_{{\cal N}=2} & = & 
Q_+Q_-\frac{1}{2g_0^2}\sum_x\, \tr \left[-iB(x)\hat{\Phi}(x) - 
\sum_{\mu=1}^2\psi_{+\mu}(x)\psi_{-\mu}(x)-\chi_+(x)\chi_-(x) \right. \nn \\ 
 & & \hspace{2.5cm} \left. -\frac14\eta_+(x)\eta_-(x)\right], 
\label{lat_N=4_Shat}
\eea
and otherwises 
\beq
\hat{S}^{{\rm LAT}}_{{\cal N}=4} = + \infty.   
\eeq
$\hat{\Phi}(x)$ is same as (\ref{Phihat_2d}). 
Similarly to the ${\cal N}=2$ case, 
the locality of the theory is satisfied, and  
the Boltzmann weight 
$\exp\left[-\hat{S}^{{\rm LAT}}_{{\cal N}=4}\right]$ is smooth and 
differentiable.   
It leads the $Q_{\pm}$ invariance of the Boltzmann weight in the same sense 
as (\ref{Q_invariance}). 
Also, the SU$(2)_R$ and $Q_+\leftrightarrow Q_-$ symmetries 
(\ref{SU2R}, \ref{Q_exchange_2d}) are not influenced by the modification. 

Arguments about the classical continuum limit and 
the absence of fermion doubling go parallel as in the ${\cal N}=2$ case. 
The fermion kinetic terms are represented in the same form 
as (\ref{wilson_like_N=2}) 
with $\Psi^T = \left(\psi_{+1}, \psi_{+2}, \chi_+, \frac12\eta_+, 
\psi_{-1}, \psi_{-2}, \chi_-, \frac12\eta_-\right)$ 
and 
\bea
 & & \gamma_1 = -i\left(\begin{array}{cccc} 
        &         &         & \sigma_1 \\
        &         & -i\sigma_2 &       \\
        & i\sigma_2 &       &          \\
\sigma_1 &        &         &          \end{array}\right), \qquad 
\gamma_2 = -i\left(\begin{array}{cccc} 
        &         &         & -\sigma_3 \\
        &         & {\bf 1}_2 &       \\
        & {\bf 1}_2 &       &          \\
-\sigma_3 &        &         &          \end{array}\right), \nn \\      
 & & P_1 = -i\left(\begin{array}{cccc} 
        &         &          & i\sigma_2 \\
        &         & -\sigma_1 &         \\
        & \sigma_1 &       &          \\
 i\sigma_2 &         &         &          \end{array}\right), \qquad  
 P_2 = -i\left(\begin{array}{cccc} 
        &         &          & {\bf 1}_2 \\
        &         & \sigma_3 &         \\
        & -\sigma_3 &       &          \\
 -{\bf 1}_2 &         &         &          \end{array}\right). 
\label{gamma_P_N=4} 
\eea
The matrices satisfy the anticommuting relation 
(\ref{anticommute_N=2}) again, which leads (\ref{D_square}) indicating 
nonexistence of fermion doublers. 

As for the renormalization in the SU$(N)$ case, 
the gauge invariance and   
the SU$(2)_R$ symmetry allow the operators $\tr(4\phi\bar{\phi} + C^2)$ and 
$\tr B^2$, 
but they are not admissible from the supersymmetry $Q_{\pm}$. 
Also, for the U$(N)$ case, we should further consider the possibility of the operators 
$\tr B$, $\tr \tilde{H}_{\mu}$, $\tr H$ induced. They are forbidden by the symmetries under  
$Q_+ \leftrightarrow Q_-$ (\ref{Q_exchange_2d}) and the reflection (\ref{reflection}). 
Thus,   
no relevant or marginal operators appear through the loop effect 
except the identity operator. The continuum ${\cal N}=4$ theory is obtained 
with no tuning of parameters. 


\setcounter{equation}{0}
\section{Modification of $\Phi(x)$}
\label{sec:modification1}

In this section we consider another possibility to single out the vacuum $U_{12}(x)=1$ 
from the degeneracy specifying the case $G=\mbox{SU}(N)$.

First we try to add a term $\Delta \Phi(x)$ to $\Phi(x)$: 
\beq
\Phi(x) \rightarrow \Phi(x) + \Delta \Phi(x), \qquad 
\Delta\Phi(x) \equiv -r(2-U_{12}(x) - U_{21}(x))  
\label{modification1}
\eeq
with the parameter $r$ appropriately chosen. 
Since $H(x)$ is traceless, the classical vacua are determined by 
\beq
\Phi(x) + \Delta\Phi(x) - \left(\frac{1}{N}\tr[\Phi(x) + \Delta\Phi(x)]\right){\bf 1}_N =0. 
\label{vacuaSU(N)}
\eeq 
However, it turns out that it does not completely resolve the degeneracy. 
For instance, we can easily see that the center elements (\ref{SU(N)minima2}) are still 
the solutions for arbitrary $r$. 

On the other hand, the equation  
\beq
\Phi(x) + \Delta\Phi(x)=0  
\label{vacuaU(N)}
\eeq 
for $G=\mbox{SU}(N)$ has the unique solution $U_{12}(x)=1$ 
with appropriately chosen $r$ as explained in appendix A. 

\subsection{${\cal N}=2$ Case}

For the ${\cal N}=2$ theory, 
we extend $\chi(x)$, $H(x)$ to the hermitian matrices $\hat{\chi}(x)$ , $\hat{H}(x)$ 
with nonzero trace parts to 
introduce the variables $\chi^{(0)}(x)$, $H^{(0)}(x)$ 
proportional to the unit matrix:
\bea
 & & \chi^{(0)}(x) = \underline{\chi}^{(0)}(x) \, {\bf 1}_N, \qquad 
     H^{(0)}(x) = \underline{H}^{(0)}(x) \, {\bf 1}_N \nn \\
 & & \hat{H}(x) = H(x) + H^{(0)}(x), \qquad \hat{\chi}(x) = \chi(x) + \chi^{(0)}(x). 
\eea
The fields with the uniderline mean the coefficients proportional to the unit matrix. 
The $Q$-transformation (\ref{Q_lattice}) of $\chi(x)$ and $H(x)$ is naturally extended to 
\beq
Q\hat{\chi}(x)= \hat{H}(x), \qquad Q\hat{H}(x) = [\phi(x), \hat{\chi}(x)] 
\eeq
with 
\beq
Q\chi^{(0)}(x) = H^{(0)}(x), \qquad QH^{(0)}(x) = 0. 
\eeq

The lattice action is modified as 
\bea
S^{{\rm LAT}}_{{\cal N}=2} & = & Q\frac{1}{2g_0^2}\sum_x \, \tr\left[ 
\frac14 \eta(x)\, [\phi(x), \,\bar{\phi}(x)] 
-i\hat{\chi}(x)\left(\Phi(x) + \Delta\Phi(x)\right)
+\hat{\chi}(x)\hat{H}(x)\right. \nn \\
 & & \hspace{2cm}\left. \frac{}{} 
+i\sum_{\mu=1}^2\psi_{\mu}(x)\left(\bar{\phi}(x) - 
U_{\mu}(x)\bar{\phi}(x+\hat{\mu})U_{\mu}(x)^{\dagger}\right)\right] 
\label{U(N)lat_N=2_S}   \\
& = & \frac{1}{2g_0^2}\sum_x \, \tr\left[
\frac14 [\phi(x), \,\bar{\phi}(x)]^2 + \hat{H}(x)^2 
-i\hat{H}(x)\left(\Phi(x) + \Delta \Phi(x)\right) \right. \nn \\
 & & \hspace{1.5cm}
+\sum_{\mu=1}^2\left(\phi(x)-U_{\mu}(x)\phi(x+\hat{\mu})U_{\mu}(x)^{\dagger}
\right)\left(\bar{\phi}(x)-U_{\mu}(x)\bar{\phi}(x+\hat{\mu})
U_{\mu}(x)^{\dagger}\right) \nn \\
 & & \hspace{1.5cm} -\frac14 \eta(x)[\phi(x), \,\eta(x)] 
- \chi(x)[\phi(x), \,\chi(x)] \nn \\
 & & \hspace{1.5cm}
-\sum_{\mu=1}^2\psi_{\mu}(x)\psi_{\mu}(x)\left(\bar{\phi}(x)  + 
U_{\mu}(x)\bar{\phi}(x+\hat{\mu})U_{\mu}(x)^{\dagger}\right) \nn \\
 & & \hspace{1.5cm}   + i\hat{\chi}(x) Q\left(\Phi(x) + \Delta\Phi(x)\right)\nn \\
 & & \hspace{1.5cm}\left. \frac{}{}
-i\sum_{\mu=1}^2\psi_{\mu}(x)\left(\eta(x)-
U_{\mu}(x)\eta(x+\hat{\mu})U_{\mu}(x)^{\dagger}\right)\right]. 
\label{U(N)lat_N=2_S2}
\eea

Due to the trace part of $\hat{H}(x)$, the minimum of the gauge part is uniquely determined by 
(\ref{vacuaU(N)}) with $r = \cot \varphi$ such that 
\beq
e^{i2\ell\varphi}\neq 1 \quad \mbox{for} \quad {}^\forall \ell=1,
\cdots, N, 
\label{other_choices_text}
\eeq
as explained in appendix A. 
On the other hand, the kinetic term of $\hat{\chi}(x)$ is 
\beq
\frac{1}{2g_0^2}\sum_x \left\{\tr \left[i\chi(x)Q(\Phi(x) + \Delta\Phi(x))\right] 
+ i\underline{\chi}^{(0)}(x)Q\,\tr(\Phi(x) + \Delta\Phi(x))\right\}.   
\eeq
Since the second term in the brace is of the order $O(a^6)$, 
it vanishes in the contiuum limit and $\underline{\chi}^{(0)}(x)$ become fermion zero-modes. 
If we integrate out $\underline{\chi}^{(0)}(x)$, we will have the nontrivial constraints 
\beq
0 = Q\, \tr(\Phi(x) + \Delta\Phi(x))
\eeq
leading $0= \tr \left[F_{12}(x)(D_1\psi_2(x)-D_2\psi_1(x))\right]$ 
at the nontrivial leading order $O(a^{9/2})$ in the continuum. 
Because such constraints are not imposed in the target continuum theory, we should avoid obtaining them. 
In order to do so, we soak up the would-be fermion zero-modes in the path-integral to consider the 
measure 
\beq
\dd \mu_{{\cal N}=2} \equiv \dd \mu_{{\rm SU}(N)\, {\cal N}=2}
 \prod_x\left(\dd \underline{H}^{(0)}(x)\dd \underline{\chi}^{(0)}(x)\, \underline{\chi}^{(0)}(x)\right)  
\eeq
with $\dd \mu_{{\rm SU}(N)\, {\cal N}=2}$ being the measure for the SU$(N)$ variables. 
Note that $\dd \underline{\chi}^{(0)}(x)\, \underline{\chi}^{(0)}(x)$ is U$(1)_R$ invariant, because 
$\dd \underline{\chi}^{(0)}(x)$ transforms same as the derivative 
$\partial /\partial \underline{\chi}^{(0)}(x)$. 
However, the insertion of the would-be zero-modes violates the $Q$ invariance as 
\beq
Q\left(\dd \underline{H}^{(0)}(x)\dd \underline{\chi}^{(0)}(x)\, \underline{\chi}^{(0)}(x)\right)
= -\dd \underline{H}^{(0)}(x)\dd \underline{\chi}^{(0)}(x)\,\underline{H}^{(0)}(x), 
\label{Q_measure}
\eeq
although the action (\ref{U(N)lat_N=2_S}) is manifestly $Q$ invariant. 

Here we consider the observables consisting the operators in the SU$(N)$ sector i.e. independent of 
$H^{(0)}(x)$ and $\chi^{(0)}(x)$. Let us write the action as 
\bea
S^{{\rm LAT}}_{{\cal N}=2} & = & S^{{\rm LAT}}_{{\rm SU}(N)\, {\cal N}=2} + 
      \frac{N}{2g_0^2}\sum_x\left[\underline{H}^{(0)}(x)^2 
           -i\underline{H}^{(0)}(x)\frac{1}{N}\tr(\Phi(x) + \Delta\Phi(x)) \right. \nn \\
            & & \hspace{3cm}\left. +i\underline{\chi}^{(0)}(x)Q\frac{1}{N}\tr (\Phi(x) + \Delta\Phi(x))\right], 
\label{action_bunri}\\
S^{{\rm LAT}}_{{\rm SU}(N)\, {\cal N}=2} & = &    Q\frac{1}{2g_0^2}\sum_x \, \tr\left[ 
\frac14 \eta(x)\, [\phi(x), \,\bar{\phi}(x)] 
-i\chi(x)\left(\Phi(x) + \Delta\Phi(x)\right)
+\chi(x)H(x)\right. \nn \\
 & & \hspace{2cm}\left. \frac{}{} 
+i\sum_{\mu=1}^2\psi_{\mu}(x)\left(\bar{\phi}(x) - 
U_{\mu}(x)\bar{\phi}(x+\hat{\mu})U_{\mu}(x)^{\dagger}\right)\right], 
\eea
so that the dependence of $\underline{H}^{(0)}(x)$ and $\underline{\chi}^{(0)}(x)$ can be explicitly seen. 
{}From (\ref{Q_measure}), (\ref{action_bunri}), 
the $Q$-transformation of $\dd \mu_{{\cal N}=2}\,e^{-S^{{\rm LAT}}_{{\cal N}=2}}$ leads 
\bea
 & & \int Q\left(\dd \mu_{{\cal N}=2}\,e^{-S^{{\rm LAT}}_{{\cal N}=2}}\right) = 
  \int \dd \mu_{{\rm SU}(N)\, {\cal N}=2}\, e^{-S^{{\rm LAT}}_{{\rm SU}(N)\, {\cal N}=2}}\, 
   \left(\prod_x\dd \underline{H}^{(0)}(x)\right) \nn \\
  &  & \hspace{2cm}\times \sum_x \left[\frac{i}{2g_0^2} \underline{H}^{(0)}(x)Q\,\tr(\Phi(x) + \Delta\Phi(x))\right] \nn \\
 & & \hspace{2cm}\times \exp\left\{-\frac{N}{2g_0^2}\sum_x \left[ \underline{H}^{(0)}(x)^2 
            -i \underline{H}^{(0)}(x)\frac{1}{N}\tr(\Phi(x) + \Delta\Phi(x))\right]\right\}
\eea
after integrating out $\underline{\chi}^{(0)}$. As the result of the integration of $\underline{H}^{(0)}$, we obtain 
\bea
\int Q\left(\dd \mu_{{\cal N}=2}\,e^{-S^{{\rm LAT}}_{{\cal N}=2}}\right) & = & 
  \int \dd \mu_{{\rm SU}(N)\, {\cal N}=2}\, e^{-S^{{\rm LAT}}_{{\rm SU}(N)\, {\cal N}=2}}\, 
   e^{-\frac{1}{8Ng_0^2}\sum_x\left[\tr(\Phi(x) + \Delta\Phi(x))\right]^2} \nn \\
 & & \times Q\sum_x\frac{-1}{8Ng_0^2}\left[\tr(\Phi(x) + \Delta\Phi(x))\right]^2,  
\eea
which means that the insertion of the would-be fermion zero-modes is equivalent to adding the supersymmetry breaking term 
\beq
\Delta S = \frac{1}{8Ng_0^2}\sum_x\left[\tr(\Phi(x) + \Delta\Phi(x))\right]^2
\label{Delta_S}
\eeq
to the $Q$ invariant action $S^{{\rm LAT}}_{{\rm SU}(N)\, {\cal N}=2}$. 
$\Delta S$ supplies the trace part of $\Phi(x) + \Delta\Phi(x)$ leading the condition for the minima (\ref{vacuaSU(N)}) 
to resolve the degeneracy. In the continuum limit, $\Delta S$ is of the order $O(a^4)$: 
$\Delta S = \frac{a^4}{8Ng^2}\int \dd^2x\, \left(\tr F_{12}(x)^2\right)^2$ and becomes irrelevant. 
Comparing to the supersymmetry breaking term (\ref{SUSY_breaking}) of the order $O(a^s)$ $(0<s<2)$ 
introduced in \cite{sugino}, (\ref{Delta_S}) becomes much more irrelevant in the continuum.

\paragraph{No Fermion Doublers and Renormalization}

The action $S^{{\rm LAT}}_{{\rm SU}(N)\, {\cal N}=2}+ \Delta S$ with $r=\cot\varphi$ 
satisfying (\ref{other_choices_text}) has the unique minimum $U_{12}(x)=1$, 
which justifies expanding the exponential of the link variables 
(\ref{unitary}). 
Note that the modification $\Delta\Phi(x)$ and $\Delta S$ does not affect the classical 
continuum limit (\ref{ccl}). 
Compared with $\Phi(x)$ giving $O(a^2)$ contributions, 
$\Delta\Phi(x)$ is of the negligible order $O(a^4)$. 
Furthermore, it turns out that  
the shift of the fermionic part of the action 
\beq
\tr[i\chi(x) Q\Delta\Phi(x)] = 
\tr\left[ir\chi(x)\left(QU_{12}(x) + QU_{21}(x)\right)\right]
\eeq  
leads to gauge-fermion interaction terms of the irrelevant order $O(a^5)$, 
but not to fermion kinetic terms.    
Thus, the fermion bilinear kinetic terms are not influenced by 
the modification, and   
no fermion doublers appear from the same argument as in the 
previous paper \cite{sugino}. 
Also, since the supersymmetry breaking effect comes in via the vertices of $\Delta S$ in the loop expansion, 
it becomes irrelevant in the continuum limit as discussed in the previous paper \cite{sugino}. 
The renormalization argument goes 
parallel to the previous paper to show that the target continuum theory 
is obtained without any fine-tuning.

\subsection{${\cal N}=4$ Case} 

For the ${\cal N}=4$ case, we can repeat the similar argument to the ${\cal N}=2$ model.
We introduce the degrees of the freedom of the trace part for $B(x)$, $\chi_{\pm}(x)$ and $H(x)$ as 
\bea
 & & \hspace{-7mm} 
  B^{(0)}(x) = \underline{B}^{(0)}(x) \, {\bf 1}_N, \qquad \chi_\pm^{(0)}(x) = \underline{\chi_\pm}^{(0)}(x)\, {\bf 1}_N, 
     \qquad H^{(0)}(x) = \underline{H}^{(0)}(x)\, {\bf 1}_N, \nn \\
 & & \hspace{-7mm} \hat{B}(x) = B(x) + B^{(0)}(x), \quad \hat{\chi_{\pm}}(x) = \chi_{\pm}(x) + \chi_{\pm}^{(0)}(x), \quad 
        \hat{H}(x) = H(x) + H^{(0)}(x). 
\label{tracepart_ari}
\eea
Again, the fields with the superscript `(0)' proportional to the unit matrix, 
and their proportional coefficients are denoted by putting the underline. 
Defined as 
\bea
 & & Q_\pm B^{(0)}(x) = \chi_\pm^{(0)}(x), \qquad Q_\pm \chi_\pm^{(0)}(x) = 0, 
       \qquad Q_\mp\chi_\pm^{(0)}(x) = \mp H^{(0)}(x), \nn \\
 & & Q_\pm H^{(0)}(x) = 0,         
\eea
the transformation rule (\ref{group_B_2d}) is naturally extended to the variables with the trace parts 
(\ref{tracepart_ari}). 

Let us consider the lattice action 
\bea
S^{{\rm LAT}}_{{\cal N}=4} & = &  
Q_+Q_-\frac{1}{2g_0^2}\sum_x\, \tr \left[-i\hat{B}(x)\left(\Phi(x)+ \Delta\Phi(x)\right) - 
\sum_{\mu=1}^2\psi_{+\mu}(x)\psi_{-\mu}(x)-\hat{\chi}_+(x)\hat{\chi}_-(x) \right. \nn \\ 
 & & \hspace{2.5cm} \left. -\frac14\eta_+(x)\eta_-(x)\right]. 
\label{U(N)lat_N=4_S}
\eea                     
{}From the argument parallel to the ${\cal N}=2$ case, we observe that $\underline{\chi_{\pm}}^{(0)}(x)$ and 
$\underline{B}^{(0)}(x)$ become the zero-modes in the continuum limit. After integrating out those, we will obtain 
the nontrivial constraints
\beq
0= Q_+Q_-\tr(\Phi(x)+ \Delta\Phi(x)), \qquad 0= Q_\pm\tr(\Phi(x)+ \Delta\Phi(x)),  
\eeq
meaning $0 = \tr\left[F_{12}(x)\left(D_1\psi_{\pm 2}(x)-D_2\psi_{\pm 1}(x)\right)\right]$ in the continuum, 
which do not appear in the target continuum theory. 
In order to avoid the constraints, we soak up the would-be zero-modes to consider the path-integral measure 
\bea
& & \dd \mu_{{\cal N}=4} \equiv  \dd \mu_{{\rm SU}(N)\, {\cal N}=4} \nn \\
 & & \quad \times\prod_x\left(\dd \underline{H}^{(0)}(x)\dd \underline{B}^{(0)}(x) \delta\left(\underline{B}^{(0)}(x)\right) 
 \dd \underline{\chi_+}^{(0)}(x)\, \underline{\chi_+}^{(0)}(x)
 \dd \underline{\chi_-}^{(0)}(x)\, \underline{\chi_-}^{(0)}(x)\right)   
\eea 
with $\dd \mu_{{\rm SU}(N)\, {\cal N}=4}$ being the measure with respect to the variables in the SU$(N)$ sector. 
The measure $\dd \mu_{{\cal N}=4}$ is invariant under the SU$(2)_R$ rotation, but not under $Q_\pm$ due to the insertion 
of the would-be zero-modes $\underline{\chi_\pm}^{(0)}(x)$, 
although $S^{{\rm LAT}}_{{\cal N}=4}$ is manifestly supersymmetric. 
Following the similar procedure to the ${\cal N}=2$ case, we end up with 
\bea
\int Q_\pm\left(\dd \mu_{{\cal N}=4}\,e^{-S^{{\rm LAT}}_{{\cal N}=4}}\right) & = & 
  \int \dd \mu_{{\rm SU}(N)\, {\cal N}=4}\, e^{-S^{{\rm LAT}}_{{\rm SU}(N)\, {\cal N}=4}}\, 
   e^{-\frac{1}{8Ng_0^2}\sum_x\left[\tr(\Phi(x) + \Delta\Phi(x))\right]^2} \nn \\
 & & \times Q_\pm\sum_x\frac{-1}{8Ng_0^2}\left[\tr(\Phi(x) + \Delta\Phi(x))\right]^2,  
\eea
again showing that the soak-up of the would-be zero-modes is equivalent to adding the supersymmetry breaking term 
(\ref{Delta_S}). $S^{{\rm LAT}}_{{\rm SU}(N)\, {\cal N}=4}$ is nothing but the action (\ref{lat_N=4_S_2d}) 
with the replacement (\ref{modification1}) made.  

Similarly to the previous cases, it is shown that the fermion doublers do not appear and 
that the target continuum theory is obtained without any fine-tuning.

\setcounter{equation}{0}
\section{Summary and Discussions}
\label{sec:summary}

In this paper, two-dimensional $G=\mbox{U}(N)$, $\mbox{SU}(N)$ super Yang-Mills theories with 
${\cal N}=2, \,4$  
supersymmetries have been constructed on the square lattice, 
keeping one or two supercharges exactly. 
We have resolved a problem of the degenerate classical vacua, 
which was encountered in the previous paper \cite{sugino}, 
{\em with keeping the exact supersymmetry}. 
Thus, any supersymmetry breaking terms do not need to be introduced, and 
the formulations exactly realize the supersymmetries 
$Q$ or $Q_{\pm}$ at the lattice level. 
Our lattice models define the continuum SYM theories without any fine-tuning. 

We have 
considered two different kinds of modifications of the actions to resolve 
the problem. 
One is a modification to impose the admissibility condition on each plaquette 
variable $U_{12}(x)$ with changing the action somewhat analogous to  
ref. \cite{luscher}. It will be also applicable to other gauge groups with an 
appropriate choice of $\epsilon$.   
It would be worth while to pursue that direction to discuss 
topological structures of the space of the admissible 
lattice gauge fields as in \cite{luscher}. 
The other modification is a simple one to 
add the term $\Delta\Phi(x)$ to the function $\Phi(x)$ with the trace parts introduced for 
some adjoint fields in $G=\mbox{SU}(N)$. 
The actions are supersymmetric, but they contain would-be zero modes which induce nontrivial constraints 
not seen in the target continuum theories. In order to avoid getting 
the constraints we have soaked up the zero-modes, 
which leads breaking of the supersymmetry. The breaking effect is irrelevant in the continuum limit, and 
it can be shown that the target continuum theories are obtained without any fine-tuning.   

It is interesting to consider appropriate modifications to four-dimensional 
models with 
\beq
\Phi_{\bA}(x) =  -i\left[U_{{\bA} 4}(x) -U_{4\bA}(x) 
+ \frac12\sum_{{\bB},{\bC}=1}^3\varepsilon_{\bA\bB\bC}\,
(U_{\bB\bC}(x) - U_{\bC\bB}(x))\right],  
\eeq
which correspond to ${\cal N}=2, \, 4$ theories \cite{sugino2}. 
Even if suitable modifications are found and exact supersymmetries are 
realized at the lattice level, 
generically four-dimensional models would need some fine-tuning of 
parameters to define the desired continuum limit. 
For the ${\cal N}=4$ case, however the situation might seem to be 
subtle, 
because it is believed to receive no quantum corrections in any order of
the perturbation theory (for example, see \cite{mandelstam}). 
It would be intriguing to compute radiative 
corrections in the lattice perturbation theory for the modified ${\cal N}=4$ 
model with the exact $Q_{\pm}$ supersymmetry.   
%

Since the same kind of degeneracy problem exists in models 
proposed in ref. \cite{elitzur}, 
the methods discussed here might be useful to resolve the difficulty there.

\acknowledgments
The author would like to thank S.~Iso, Y.~Kikukawa and H.~Umetsu 
for variable discussions, 
and H.~Itoyama, Y.~Okada and H.~Suzuki for useful conversations. 
In particular, he would like to express his gratitude to Y. Kikukawa 
for pointing out the problem in the $G=\mbox{SU}(N)$ case in the previous version 
of the paper. 
He also thank the theory group members of KEK and RIKEN  
for their warm hospitality during his stay.


\appendix
\section{Uniqueness of the solution of eq. (\ref{vacuaU(N)})}
\label{sec:appendix}
\setcounter{equation}{0}
\renewcommand{\theequation}{A.\arabic{equation}}

Here we show that the equation (\ref{vacuaU(N)}) for $G=\mbox{SU}(N)$ 
has a unique solution 
$U_{12}(x)=1$ for appropriate choices for the parameter $r$. 

First we shall take as 
\beq
r=\cot \frac{\pi}{2N}. 
\label{choice_r}
\eeq 
For other choices, we will discuss later. 
{}From the equation (\ref{vacuaU(N)}), $U_{12}(x)$ can be expressed as the diagonalized form  
\beq
U_{12}(x) = \Omega(x) \left(\begin{array}{ccc} e^{i\theta_1(x)} &  &  \\
  &  \ddots &  \\  &  & e^{i\theta_N(x)}\end{array}\right) \Omega(x)^\dagger 
\label{diagonalize_U12}
\eeq
with $\Omega(x) \in \mbox{SU}(N)$, 
$\theta_N(x) \equiv -\theta_1(x) - \cdots -\theta_{N-1}(x)$, 
and $-\pi < \theta_i(x) \leq +\pi$ ($i=1, \cdots, N-1$). 
Under this parameterization, 
the equation for the minima (\ref{vacuaU(N)}) reads 
\bea
\sin \theta_1(x) & =  & 2r \sin^2\left(\frac{\theta_1(x)}{2}\right), \nn \\
                 & \vdots &                           \nn \\
\sin \theta_{N-1}(x) & =  & 2r \sin^2\left(\frac{\theta_{N-1}(x)}{2}\right), 
\nn \\  
-\sin(\theta_1(x) + \cdots + \theta_{N-1}(x)) & = & 
2r \sin^2\left(\frac{\theta_1(x) + \cdots + \theta_{N-1}(x)}{2}\right). 
\label{vacua_theta}
\eea
With the choice (\ref{choice_r}), 
the first $N-1$ equations give the solutions 
$\theta_i(x) = 0$ or $\frac{\pi}{N}$ ($i=1, \cdots, N-1$). 
Among them, 
the last equation allows the only one combination   
\beq
(\theta_1(x), \cdots, \theta_{N-1}(x))=(0, \cdots, 0), 
\label{unique_sol}
\eeq
which is nothing but $U_{12}(x)=1$. 
(Any other combinations of the solutions 
are not compatible to 
the last equation, because the LHS is evaluated to be negative 
while the RHS positive.)  

Let us discuss about other choices of $r$. Parameterizing as $r=\cot\varphi$, 
\beq
\Phi(x)+\Delta\Phi(x) = \frac{-1}{\sin\varphi}\left[
e^{-i\varphi}(1-U_{12}(x)) + e^{i\varphi}(1-U_{21}(x))\right], 
\eeq
where $\Phi(x)+\Delta\Phi(x)=0$ means that $e^{-i\varphi}(1-U_{12}(x))$ 
is anti-hermitian. Configurations of $U_{12}(x)$ giving the minima are 
expressed as 
\beq
U_{12}(x)=1-ie^{i\varphi}T(x)
\label{otherU}
\eeq
with $T(x)$ being hermitian.   
The eigenvalues of $T(x)$ are denoted as $t_i(x)$ ($i=1, \cdots, N$). 
The unitary condition $U_{12}(x)U_{12}(x)^\dagger=1$ determines $t_i(x)$ as 
\beq
t_i(x)= 0 \quad \mbox{or}\quad -2\sin\varphi \qquad \mbox{for}\quad 
i=1, \cdots, N. 
\eeq
The unitarity alone does not uniquely fix $t_i(x)$. 
However, the unimodular condition 
\beq
\det U_{12}(x)=1 
\label{unimodular}
\eeq
gives further constraints. 
For the case that $\ell$ eigenvalues of $t_i(x)$ are $-2\sin\varphi$ and 
the remaining $N-\ell$ are 0, (\ref{unimodular}) leads 
\beq
e^{i2\ell\varphi}=1. 
\eeq
Thus, taking $\varphi$ such that  
\beq
e^{i2\ell\varphi}\neq 1 \quad \mbox{for} \quad {}^\forall \ell=1,
\cdots, N, 
\label{other_choices}
\eeq
the equation $\Phi(x)+\Delta\Phi(x)=0$ has the unique solution 
$t_1(x)=\cdots =t_N(x)=0$, equivalent to $U_{12}(x)=1$.
Of course, the choice (\ref{choice_r}) satisfies (\ref{other_choices}).


\end{document}